\documentclass{mn2e}
\usepackage{natbib}
\usepackage{amsfonts}
\usepackage{amsmath}
\usepackage{amssymb}
\usepackage{graphicx}
\usepackage{color}

\newcommand{\pino}{{\sc Pinocchio}}

\newcommand{\p}{\ensuremath{\partial}}
\newcommand{\del}{\ensuremath{\delta}}
\newcommand{\Del}{\ensuremath{\Delta}}
\newcommand{\lam}{\ensuremath{\lambda}}

\newcommand{\sig}{\ensuremath{\sigma}}

\newcommand{\epc}{\ensuremath{\epsilon_{\times}}}
\newcommand{\Sc}{\ensuremath{S_{\times}}}
\newcommand{\So}{\ensuremath{S_{0}}}
\newcommand{\delc}{\ensuremath{\delta_{\rm c}}}
\newcommand{\delo}{\ensuremath{\delta_{0}}}
\newcommand{\delh}{\ensuremath{\delta_{\rm h}}}
\newcommand{\rhoh}{\ensuremath{\rho_{\rm h}}}

\newcommand{\Msun}{\ensuremath{M_{\odot}}}
\def\Mpc{\, h^{-1}{\rm Mpc}}

\def\Ms{h^{-1}M_{\odot}}

\newcommand{\avg}[1]{\ensuremath{\left\langle \,#1\, \right\rangle}}

\newcommand{\der}{\ensuremath{{\rm d}}}

\newcommand{\eqn}[1]{equation~\eqref{#1}}
\newcommand{\eqns}[1]{equations~\eqref{#1}}
\newcommand{\fig}[1]{Figure~\ref{#1}}
\newcommand{\figs}[1]{Figures~\ref{#1}}
\newcommand{\ph}[1]{\phantom{#1}}

\newcommand{\be}{\begin{equation}}
\newcommand{\ee}{\end{equation}}

\title[Bias deconstructed]
      {Bias deconstructed:  Unravelling the scale dependence of halo bias using real space measurements} 
\author[Paranjape et al.]
{\parbox[t]{\textwidth}{Aseem Paranjape$^{1,2}$\thanks{Email: aseemp@phys.ethz.ch}, Emiliano Sefusatti$^{2,3}$, Kwan Chuen Chan$^{4}$, Vincent Desjacques$^{4}$, Pierluigi Monaco$^{5,6}$ \& Ravi K. Sheth$^{2,7}$}\\\,\\
  $^{1}$ ETH Z\"urich, Institute for Astronomy, 
      Wolfgang-Pauli-Strasse 27, CH-8093 Z\"urich -- Switzerland\\
  $^{2}$ The Abdus Salam International Center for Theoretical Physics,
      Strada Costiera, 11, Trieste 34151 -- Italy\\
  $^{3}$ INAF, Osservatorio Astronomico di Brera, Via Bianchi 46, I-23807 Merate (LC) -- Italy\\
  $^{4}$ D\'epartement de Physique Th\'eorique and 
      Center for Astroparticle Physics (CAP), Universit\'e de Gen\`eve, \\
      $\quad$24 quai Ernest Ansermet, CH-1211 Gen\`eve -- Switzerland\\
  $^{5}$ Dipartimento di Fisica - Sezione di Astronomia, Universit\`a di Trieste, via Tiepolo 11, I-34131 Trieste -- Italy\\
  $^{6}$ INAF, Osservatorio Astronomico di Trieste, Via Tiepolo 11, I-34131 Trieste -- Italy\\
  $^{7}$ Center for Particle Cosmology, University of Pennsylvania, 
      209 S. 33rd St., Philadelphia, PA 19104 -- USA
}
\begin{document}
\pagerange{\pageref{firstpage}--\pageref{lastpage}}

\maketitle 

\label{firstpage}

\begin{abstract}
\noindent 
We explore the scale dependence of halo bias using real space cross-correlation measurements in $N$-body simulations and in \pino, an algorithm based on Lagrangian Perturbation Theory. Recent work has shown how to interpret such real space measurements in terms of $k$-dependent bias in Fourier space, and how to remove the $k$-dependence to reconstruct the $k$-\emph{independent} peak-background split halo bias parameters.  We compare our reconstruction of the linear bias, which requires no free parameters, with previous estimates from $N$-body simulations which were obtained directly in Fourier space at large scales, and find very good agreement.
Our reconstruction of the quadratic bias is similarly parameter-free, although in this case there are no previous Fourier space measurements to compare with.
Our analysis of $N$-body simulations explicitly tests the predictions of the excursion set peaks (ESP) formalism of \citet{psd13} for the scale dependence of bias; we find that the ESP predictions accurately describe our measurements. In addition, our measurements in \pino\ serve as a useful, successful consistency check between \pino\ and $N$-body simulations that is not accessible to traditional measurements.
\end{abstract}


\section{Introduction}
\label{intro}
Galaxies, and the dark matter halos they live in, cluster differently from the underlying dark matter field itself. This halo bias is expected to be nonlinear, nonlocal and stochastic, and understanding its behaviour is a prerequisite to a successful program of precision cosmology with large scale structure. While this nonlinearity, nonlocality and stochasticity of bias is measured in $N$-body simulations of cold dark matter, its precise physical origin remains unclear, and is likely to be influenced by several effects \citep*{dcss10,css12,bsdm12,scs13}. In practice, in the absence of accurate analytical predictions of the so-called nonlinear bias parameters $b_n$ discussed below, one resorts to fitting these parameters to measurements in $N$-body simulations \citep*{t+05,psp12} or marginalising over them when analysing data from galaxy surveys \citep[e.g.,][]{b+11,s+12}, leading to a potential source of unmodelled systematic effects when attempting to recover information on cosmological parameters.

The language used when discussing halo bias is also not unique. Traditional measurements of bias in simulations are performed in Fourier space. For example, ``linear bias''  is typically defined using ratios of power spectra of the halo overdensity $\delh(\vec{k})$ and matter overdensity $\del(\vec{k})$. E.g., 
\be
b_1^2(k) \equiv \frac{P_{\rm hh}(k)}{P_{\rm mm}(k)} \quad \textrm{or} \quad b_1(k) \equiv \frac{P_{\rm hm}(k)}{P_{\rm mm}(k)}\,,
\label{b1Lag-Fourier}
\ee
where $P_{\rm hh}(k) = \avg{\delh^2}$, $P_{\rm mm}(k) = \avg{\del^2}$ are halo and matter auto-power spectra, respectively, and $P_{\rm hm}(k) = \avg{\delh\del}$ is the corresponding cross-power spectrum. These ratios are found to be scale-independent at large scales (small $k$) as expected from peak-background split arguments \citep{k84,mw96,st99}. 

Quadratic bias is typically estimated by measuring (cross-)bispectra of $\delh(\vec{k})$ and $\del(\vec{k})$ and modelling them, e.g., by using perturbation theory or halo model arguments combined with a ``local biasing'' scheme $\delh(\vec{x})=b_1\del(\vec{x})+b_2\del(\vec{x})^2/2!+\ldots$ \citep{fg93}, and in this case the state-of-the-art \citep{psp12} shows systematic effects associated with, e.g., shot-noise modelling. 

The corresponding real space measurements of bias typically involve gridding the halo and matter density fields on some smoothing scale and then fitting a quadratic relation to the associated scatter plot \citep[e.g.,][]{mg11}. The resulting fits show a dependence on smoothing scale, although it is not easy to interpret this scale dependence in terms of a $k$-dependence in Fourier space \citep{cs12}. \citep*[See also][for a large-scale real-space treatment that assumes local biasing and recovers scale-independent bias parameters to fourth order.]{abl08}

Recent work using an excursion set approach to the problem has revealed several features of halo bias: (a) Lagrangian Fourier-space bias at any nonlinear order is very naturally linked to a particular real-space definition of Lagrangian bias based on cross-correlating the halo density with a suitable transform of the smoothed initial matter density \citep*{ps12a,mps12}; (b) the excursion set analysis, as well as its extension to peaks theory \citep*{ps12b,psd13}, predicts a specific smoothing-scale dependence of these real-space bias parameters, and hence a specific $k$-dependence in Fourier space; and (c) this scale dependence can be unravelled to reconstruct the large scale, scale-\emph{independent} bias coefficients using measurements at a finite intermediate smoothing scale. Ultimately, it is the dependence of these scale-independent coefficients on redshift and halo mass that probes the underlying cosmology.

In this paper we apply these ideas to halos identified in $N$-body simulations of cold dark matter, as well as halos identified in {\pino} \citep*{mo02, mo13}, which is a fast algorithm based on Lagrangian Perturbation Theory which provides positions, velocities and merger histories of dark matter halos. The measurements in the $N$-body simulations constitute a direct test of the excursion set peaks (ESP) formalism \citep{psd13} which, as we show below, fares very well. The additional measurements in {\pino} then become a very useful (and successful) consistency check between {\pino} and the $N$-body simulations on the one hand, and between {\pino} and ESP on the other. Taken together, our results constitute a self-consistent test of real-space measurements of linear and quadratic bias with \emph{no} free parameters.

The paper is organised as follows. In section~\ref{sims} we give details of our simulation data set, including a brief description of the {\pino} algorithm. Section~\ref{bias} deals with measuring the bias parameters and comparing with theory. We first recapitulate in Section \ref{bias-analytical} the real-space definition of the $n^{\rm th}$ order bias parameters $b_n$ and its relation to Fourier-space definitions such as \eqn{b1Lag-Fourier}. Our measurements and the resulting estimates of $b_1$ and $b_2$ from the data are described in Section~\ref{b1b2}. We find that these estimates, which we make at different smoothing scales, are in good agreement with the corresponding (scale-dependent) predictions of the ESP formalism. This agreement is important because the measurements themselves are completely independent of the ESP formalism. 

In section~\ref{b10b20} we use the reconstruction algorithm mentioned above, specifically the version described by \citet{psd13}, to obtain estimates of the scale-\emph{independent} peak-background split parameters $b_{10}$ and $b_{20}$ from the estimates of $b_1$ and $b_2$. 
The peak-background split bias parameters defined in the excursion set approach directly probe the halo mass function $f(\delc;m)$ through
\be
b_{n0} = f^{-1}\left(-\frac{\p}{\p\delc}\right)^nf,
\label{pbg}
\ee
where \delc\ is the usual overdensity threshold predicted by spherical collapse \citep{mw96,st99}.

If the reconstruction works well, then these estimates of $b_{n0}$ should be independent of the smoothing scale at which the $b_n$ were measured; we find that this is indeed the case. The linear bias coefficient $b_{10}$ is also directly probed by the large scale limit of Fourier-space measurements such as \eqn{b1Lag-Fourier} \citep{ps12a}. We compare our reconstruction of $b_{10}$ with the Fourier-space large scale fit to $N$-body simulations provided by \citet{t+10}, and find very good agreement.
For $b_{20}$ there are no previous $N$-body measurements we can compare with; a comparison with the ESP prediction \citep{psd13} shows good agreement. We conclude in section~\ref{discuss}. 

We assume a flat $\Lambda$-cold dark matter cosmology with Gaussian initial conditions and compute transfer functions using {\sc Camb} \citep*{camb}\footnote{   http://lambda.gsfc.nasa.gov/toolbox/tb\_camb\_form.cfm} for two different sets of parameter values: $(\Omega_m,\sig_8,n_s,h,\Omega_b) = (0.272,0.81,0.967,0.704,0.0455)$ for the $N$-body simulations and $(0.25,0.8,0.95,0.7,0.044)$ for {\pino}.

\section{Simulations}
\label{sims}

\subsection{$N$-body}
\label{Nbody}

Our cold dark matter simulations were run with $1024^3$ particles in a cubic box of size $1500\Mpc$, with each particle carrying a mass of $2.37 \times 10^{11}\Ms$. Gaussian initial conditions were set at a starting redshift $z=99$, with initial particle displacements implemented using $2^{\rm nd}$ order Lagrangian Perturbation Theory \citep*{cpr06}. The simulations were run using {\sc Gadget II} \citep{gadget05}. We use six realizations, with halos identified using the Spherical Overdensity (SO) halo finder AHF \citep*{gkg04,kk09} which uses a redshift-dependent overdensity criterion motivated by spherical collapse \citep*{ecf96,bn98}. 
We only study halos having at least $100$ particles:  this corresponds to halo masses larger than $\sim10^{13.4}\Ms$. 
\begin{figure*}
 \centering
 \includegraphics[width=0.9\textwidth]{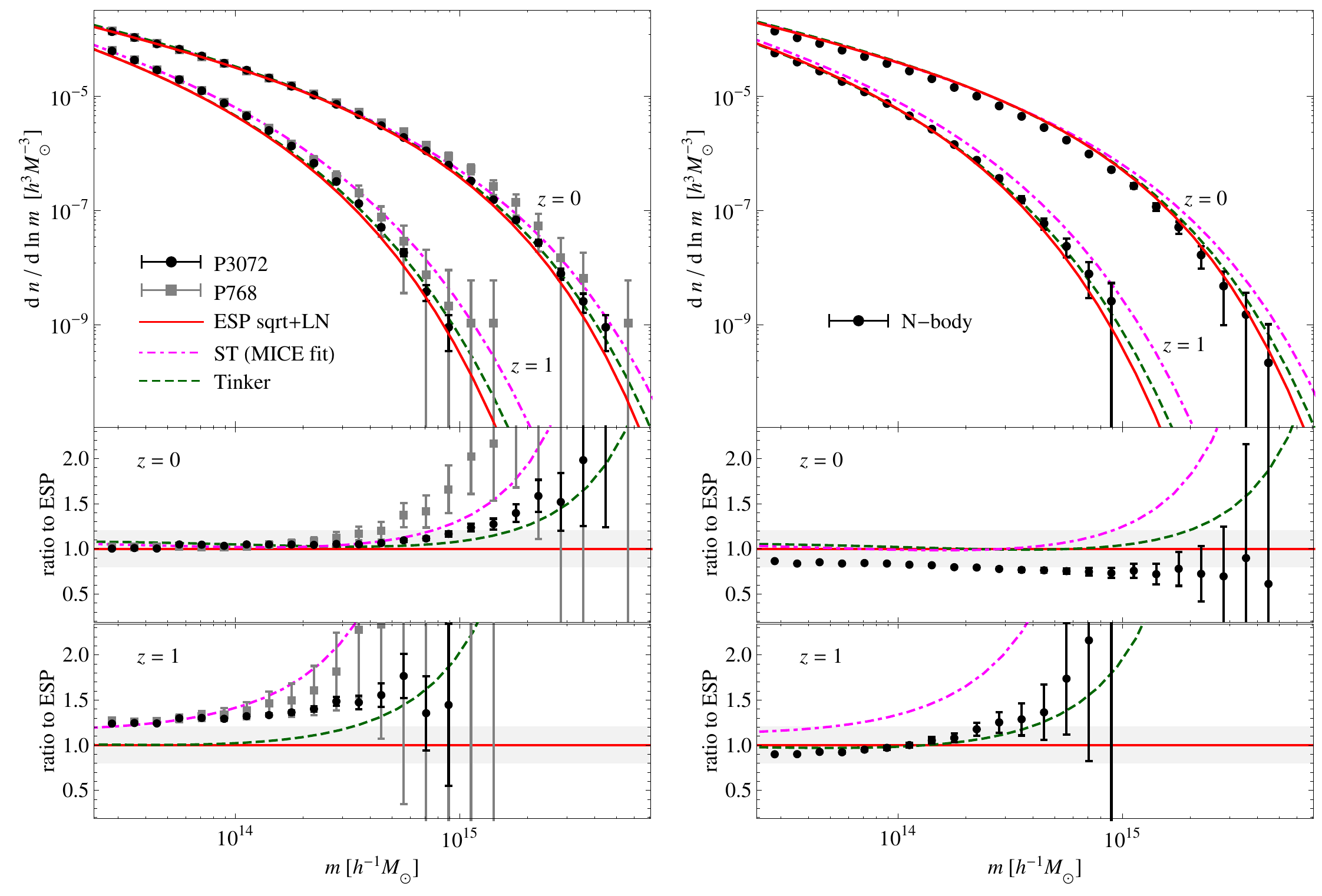}
 \caption{\emph{Left:} Halo mass functions measured in {\pino} using one realization of the $3072\Mpc$ box (black circles with error bars) and averaged over nine realizations of the $768\Mpc$ box (gray squares) at redshifts $z=0$ and $z=1$. For comparison, the smooth curves show the excursion set peaks (ESP) calculation from \cite{psd13} (solid red) and two fits to $N$-body simulations: the fit by \citet[][dashed green]{t+08} and the \citet{st99} form with their $q=0.7$ and $p=0.26$ which we find gives a good fit to the MICE mass function from \citet[][dot-dashed magenta]{mice}. \emph{Right:} Halo mass functions in our $N$-body simulations, averaged over six realizations of a $1500\Mpc$ box (black circles with error bars). The solid curves are the same as in the left panel, but computed using cosmological parameters appropriate for this $N$-body simulation. See text for a discussion. Lower panels in each case show the ratio of each quantity to the ESP predictions, with the gray band indicating $20\%$ deviations.
 }
 \label{fig-mf}
\end{figure*}

\subsection{{\large {\pino}}}
\label{pinocchio}
For a detailed explanation of the {\pino} code we refer the reader to the original paper and to the more recent \citet*{mo13} where its parallel implementation and its application to cosmological volumes are presented. Here we limit ourselves to a quite succinct description.

\pino\ starts from a linear density field generated on a grid in a manner close to the generation of initial conditions in an $N$-body simulation. It
uses 3rd-order LPT applied to the evolution of a homogeneous ellipsoid
to compute the collapse times of ``particles'' (grid points);
consistently with the excursion set approach, collapse times are
computed for many smoothing radii, thus constructing for each particle
a ``trajectory'' in the plane defined by mass variance and
inverse collapse time (the inverse of the growth factor at collapse).
At variance with the standard excursion set approach, 
correlations between trajectories of nearby particles are fully taken
into account.  
An algorithm that mimics the hierarchical assembly of
structures is then applied to construct dark matter halos. These are
built in the Lagrangian space, then displaced to their final positions
using the Zel'dovich approximation.  This algorithm has been tuned to
reproduce, to within $5$-$10\%$, the ``universal'' mass function of
\cite{wa06}.

As mentioned above, {\pino} borrows ideas
from excursion set theory 
and, on its own, cannot be considered a completely independent check
for the ESP formalism. Nevertheless, these approaches are different
enough from each other and from a full-fledged $N$-body simulation
that we believe our analysis provides valuable information on the
extent to which each of them reliably captures the underlying physical
processes.

We consider here two different set-ups for the {\pino} realizations\footnote{The two {\pino} runs correspond in size and resolution to the MICE768 and MICE3072 runs of the MICE suite \citet*{mice}. This choice is motivated by a direct comparison with those N-body simulations \citep{mo13}}. The first consists of a box of side $768\Mpc$ sampled by $1024^3$ particles with mass $2.9\times10^{10}\Ms$. The second is given instead by a larger box of side $3072\Mpc$ sampled by $2048^3$ particles, each of mass $2.3\times10^{11}\Ms$. For the smaller box we produce nine realizations with different initial seeds, in order to provide a more solid estimate of the uncertianties on our results. 
In each case we only study halos with at least 100 particles.

\citet*{mo13} show, using the same realizations, that {\pino} predicts a mass function consistent, within 10\%, to the fit of \citet{wa06} for the mass range $10^{13}$ to $10^{15}\Ms$, while a larger discrepancy is observed, in the large mass tail, with the results of \citet*{mice}.
In addition, for halo populations characterized by a fixed threshold in mass, a 10\% agreement is achieved for the linear bias, determined from measurements of the {halo-}halo power spectrum. Better results for bias are found at lower masses, where the mass function fit
is more accurate.  

\fig{fig-mf} shows the halo mass functions output in these {\pino} realizations (left panels) and measured in our $N$-body simulations (right panels) at two different redshifts, compared with the fit to $N$-body simulations by \citet[][specifically, using their SO-200mean parameter values]{t+10}, the \citet[][ST]{st99} form with their $q=0.7$ and $p=0.26$ which we find gives a good fit to the MICE simulations of \cite{mice}, and the analytical ESP calculation from \citet{psd13}. 
The lower panels in \fig{fig-mf} show the ratio of all quantities with the ESP prediction. (We will use a similar format in all our Figures below.)
As discussed above, these mass functions agree at the $10$-$20\%$ level, with larger discrepancies at higher masses. We also note that the halo masses at $z=0$ in our $N$-body simulations tend to be $\sim10\%$ lower than those expected from, e.g., the SO fit by \citet{t+10}, and this shows up as a vertical offset in the ratio in the lower panel of the Figure. This could be due to the slightly different halo finding criteria in these simulations; the \citet{t+10} fit we use corresponds to halos identified by them at a density equal to 200 times the mean density at all redshifts, while the halo finder in our simulation finds halos at $z=0$ with density $\sim350$ times the mean density and at $z=1$ with $\sim200$ times the mean density \citep{ecf96,bn98}.

\section{Halo bias in real space}
\label{bias}

\subsection{Analytical motivation}
\label{bias-analytical}

Before presenting our measurements, we summarize our current understanding of halo bias as motivated by the excursion set approach and recapitulate how the reconstruction argument works. This will also serve to set our notation. 

Throughout this paper we will focus on Lagrangian bias defined with respect to the linearly extrapolated initial dark matter density field. We will also denote by $\delc$ the traditional spherical collapse barrier for the excursion set random walks, using\footnote{
The value 1.686 strictly holds only in an $\Omega_{\rm m}=1$ cosmology -- we use it primarily for ease of comparison with \citet{psd13} and \citet{t+10} who also used 1.686. The exact value appropriate for the $\Lambda$CDM cosmologies we study would be different from this number at the $\sim1\%$ level. This would lead to discrepancies in the mass function, at the highest masses, of order $15$-$20\%$, whereas in the bias it would cause discrepancies of order $\sim1\%$ which have no impact on our final results.
} $\delc(z=0)=1.686$, and define $\nu\equiv \delc/\sqrt{s}$ where 
\be
s \equiv \sig_0^2(R) \equiv \avg{\del(R)^2} =  \int\der\ln k\,\Delta^2(k) W(kR)^2
\label{s-def}
\ee 
is the linearly extrapolated variance on the Lagrangian scale of the halo, with $\Delta^2(k)\equiv k^3P(k)/2\pi^2$ the dimensionless linear matter power spectrum at $z=0$ and $W(kR)$ the smoothing filter which we will take to be a TopHat in real space so that $W(y)=(3/y^3)(\sin y - y\,\cos y)$.

Halo bias can be defined in Fourier or real space. E.g., linear bias in Fourier space can be defined using ratios of (cross-)power spectra as in \eqn{b1Lag-Fourier}.
While these are convenient definitions, e.g., in an $N$-body simulation, analytical approaches such as the excursion set formalism work naturally in \emph{real} space, and \emph{a priori} it is not obvious how the results of the latter should be interpreted in Fourier space.
Recent work \citep{ps12a,ms12,mps12} has shown how this connection can be made in practice. In particular, \citet{mps12} argued that a useful definition of the $n^{\rm th}$ order Lagrangian halo bias coefficient $b_n$ in real space is as follows. Consider a simulation in which we have identified halos at some redshift, e.g. $z=0$.
One can now use the particles identified as belonging to a halo in the final nonlinear field to define a ``protohalo'' in the initial conditions, and compute the center-of-mass of this Lagrangian protohalo using the positions of those same particles in the initial conditions.
If there are $N$ halos (equivalently, protohalos) in a given mass bin, then the $n^{\rm th}$ order bias coefficient $b_n$ is estimated at some Lagrangian smoothing scale $R_0$ as
\be
\hat b_n = \So^{-n/2}\frac1N\sum_{i=1}^{N}H_n(\del_{0i}/\sqrt{\So})\,,
\label{bn-msd}
\ee
where $\del_{0i}$ is the \emph{dark matter} density contrast in the initial conditions (linearly extrapolated to present epoch) smoothed on the scale $R_0$ and centered on the center-of-mass in the initial conditions of the $i^{\rm th}$ halo in the bin; \So\ is the linearly extrapolated variance at scale $R_0$, 
\be
\So = \avg{\delo^2} = \sig_0^2(R_0)\,,
\label{S0}
\ee
and the $H_n$ are the probabilist's Hermite polynomials, $H_n(x)={\rm e}^{x^2/2}(-d/dx)^n {\rm e}^{-x^2/2}$, with $p_{\rm G}(x-\mu;\sig^2)$ a Gaussian in the variable $x$ with mean $\mu$ and variance $\sig^2$, so that
\be
H_1(\delo/\sqrt{\So}) = \frac{\delo}{\sqrt{\So}} \quad\textrm{and}\quad H_2(\delo/\sqrt{\So}) = \frac{\delo^2}{\So}-1\,.
\label{H1H2}
\ee
The measurement prescription in \eqn{bn-msd} requires the Lagrangian locations of the halos and the corresponding smoothed Lagrangian dark matter overdensities, but is independent of any assumptions specific to a particular excursion set-based prescription such as, e.g., ESP. The motivation for \eqn{bn-msd} is equation~(32) of \citet{mps12} \citep[see also][]{s88}:
\begin{align}
b_n &\equiv \frac1{\So^{n/2}} \avg{\rhoh H_n(\delo/\sqrt{\So})} \notag\\
&= \int_{-\infty}^\infty\der\delo\,p_{\rm G}(\delo;\So) \avg{\rhoh|\delo,\So} H_n(\delo/\sqrt{\So})\,.
\label{bn-crosscorr}
\end{align}
In the first line, $\rhoh\equiv1+\delh$ is the normalised Lagrangian halo density field of halos of mass $m$ -- essentially, a sum over Dirac delta functions at the appropriate protohalo centers-of-mass discussed above.
This formal expression
integrates over the distribution of halo-centric $R_0$-smoothed \delo-values
\citep[see][for a discussion of why this distribution is Gaussian to a very good approximation]{mps12} weighted by the normalised mass fraction $\avg{\rhoh|\delo,\So} = f(m|\delo,\So)/f(m)$ in halos surrounded by a fixed overdensity \delo\ on scale $R_0$, which is predicted by the excursion set framework (and, of course, depends on details of the implementation of the latter).

The connection to Fourier-space bias arises as follows. The excursion set analysis \citep{mps12} as well as its ESP extension \citep{ps12b,psd13} predict the following form\footnote{The convention for notation in \eqn{bn-expand} differs from the one used in, e.g., \citet{dcss10}. The convention here was introduced by \citet{mps12} and is adapted to counting powers of \epc\ in real space which roughly correspond to powers of $k^2$ in Fourier space.} for the $b_n$:
\be
b_n =\left(\frac{\Sc}{\So}\right)^n \sum_{r=0}^n \binom{n}{r}\,b_{nr}\,\epc^r\,,  
\label{bn-expand}
\ee
where
\begin{align}
\Sc &= \int\der\ln k\,\Del^2(k)W(kR)W(kR_0)\,,\notag\\
\epc &= 2\,\der\ln\Sc/\der\ln s\,
\label{Scepc}
\end{align}
are\footnote{Strictly speaking, for ESP, \epc\ should be defined in terms of mixed spectral moments:~$\epc = (s/\Sc)(\sig_{1m\times}^2/\sig_{1m}^2)$ where~$\sig_{1m}^2 = \int\der\ln k\,\Del^2(k)\,k^2{\rm e}^{-k^2R_{\rm G}^2/2}W(kR)\,;\sig_{1m\times}^2 = \int\der\ln k\,\Del^2(k)\,k^2{\rm e}^{-k^2R_{\rm G}^2/2}W(kR_0)$ and $R_{\rm G}$ is matched to $R$ by demanding $\avg{\del_{\rm G}\del}=s$ for the reasons discussed by \citet{psd13}. While we implement this in our analysis, we have found that using the second equation in \eqref{Scepc} leads to practically identical results.\label{fn1}}
cross-correlations between the mass overdensity field smoothed on the large scale $R_0$ and on the Lagrangian scale of the halo $R\propto m^{1/3}$, and the $b_{nr}$ are mass-dependent but scale-\emph{independent} coefficients 
whose functional form depends on details of the analysis, such as whether one uses the traditional excursion set calculations of \citet{mps12} or the ESP approach of \citet{psd13}.

Among these, the coefficients $b_{n0}$ are somewhat special because they are the logarithmic derivatives of the halo mass function $f(\delc,s)$ \citep{mps12} and coincide with the peak-background split bias parameters from \eqn{pbg}. These are the coefficients one is typically interested in measuring, since they carry information regarding the growth of large scale structure and are hence sensitive to the underlying cosmology. The other coefficients follow from the cross-correlation calculations advocated by \citet{mps12}; e.g., for $n=1,2$ in the ESP case they can be read off from equations (29) and (30) in \citet[][also see below]{psd13}.

\citet{mps12} showed how the appearance of \epc\ in the expression for real-space bias signals a $k$-dependence in Fourier-space bias. Essentially, this is because the real-space cross-correlation that defines $b_n$ can be interpreted in Fourier space by formally introducing $\rhoh(\vec{k})$ and then matching terms in Fourier space with the quantities appearing in \eqn{bn-expand}. Roughly, each power of \epc\ in real space corresponds to a power of $\der\ln W(kR)/\der\ln R$ in Fourier space. So, for example, if $W(kR)$ were a Gaussian filter $W(kR)={\rm e}^{-k^2R^2/2}$ then the real-space linear bias $b_1 = (\Sc/\So)\left(b_{10} + \epc b_{11}\right)$ would translate in Fourier space as $b_1(k) = b_{10} + (k^2s/\sig_1^2)b_{11}$ where $\sig_1^2 = \int\der\ln k\,\Del^2(k)k^2{\rm e}^{-k^2R^2}$. 
\citet{mps12} gave a formal proof that the real-space $b_n$ correspond to integrals over quantities that \citet{m11} calls ``renormalised'' Lagrangian bias coefficients in Fourier space. 

A remarkable aspect of the excursion set and ESP frameworks is that there exist linear relations between the $b_{nr}$ which allow \emph{all} of them to be written in terms of only the coefficients $b_{n0}$. E.g., for the simplified case of a constant excursion set barrier $B=\delc$, one finds \citep{mps12} 
\begin{align}
\delc b_{11} &= \nu^2 - \delc b_{10}\notag\\
\delc^2b_{21} &= \nu^2(\delc b_{10}-1) - \delc^2b_{20}\notag\\
\delc^2b_{22} &= \delc^2b_{20}+\nu^2(\nu^2-2\delc b_{10}+1)\,.
\label{bnr-mps12}
\end{align}
This means that the scale dependent coefficients $b_1$ and $b_2$ can be written as \emph{linear} combinations of $b_{10}$ and $b_{20}$. 
And upon measuring $b_1$ and $b_2$, one can read off the values of $b_{10}$ and $b_{20}$. This is the basis of the reconstruction procedure proposed by \citet{mps12}, and generalises to arbitrary order $b_{n0}$.

There is one complication, though. \citet{psd13} discussed the fact that the barrier appropriate for the excursion set random walks which determine halo masses is not deterministic, but has a mass-dependent scatter, and argued that this is a key ingredient in the analysis if one demands $\sim10\%$ accuracy when comparing the halo mass function with that measured in simulations. They showed, based on the $N$-body results of \citet{rktz09}, that a good model for this barrier is
\be
 B = \delc + \beta\sqrt{s}\,,
\label{sqrtbarrier}
\ee
where $\beta$ is a stochastic variable drawn from a Lognormal distribution whose mean and variance are fixed by the \citet{rktz09} results to be approximately $0.5$ and $0.25$, respectively. \citep*[The results for halo bias are very insensitive to the exact choice for this distribution; this is reassuring since ][suggest slightly different values for the mean and variance of this distribution.]{dts13} 

Define the functions
\be
\mu_n(\nu,\beta) = \nu^nH_n(\nu+\beta) \,,
\label{mu_n-beta}
\ee
and their averages over $\beta$,
\begin{align}
\avg{\mu_1|\nu} &= \nu\left(\nu+\avg{\beta|\nu}\right)\,,\notag\\
\avg{\mu_2|\nu} &= \avg{\mu_1^2|\nu} - \nu^2\notag\\
&=\nu^2\left(\nu^2-1+2\nu\avg{\beta|\nu}+\avg{\beta^2|\nu}\right)\,,
\label{mu1mu2-avg}
\end{align}
where
\be
\avg{\beta^j|\nu} = \frac{\int\der\beta\,p(\beta)f_{\rm ESP}(\nu|\beta)\beta^j}{\int\der\beta\,p(\beta)f_{\rm ESP}(\nu|\beta)}\,,
\label{beta-avg}
\ee
with $f_{\rm ESP}(\nu|\beta)$ given in equation~(13) of \citet{psd13}. \citep[The second line of equation~\ref{mu1mu2-avg} corrects a typo in equation 37 of][]{psd13}. Then the estimate for the reconstructed $b_{10}$ becomes \citep[equation 35 of][]{psd13},
\begin{align}
\delc \hat b_{10}(\nu) &= \frac1{(1-\epc)}\bigg[\frac{\delc \hat b_1}{(\Sc/\So)} - \epc\avg{\mu_1|\nu} \bigg]\,,
\label{dcb10-recon}
\end{align}
which is straightforward to implement since each term is either easily measurable or calculable. 

For $b_{20}$, however, they showed that the exact expression for the reconstruction involves a term $\avg{b_1\mu_1|\nu}$ which is cumbersome to keep track of (although it is in principle measurable in the simulation). They therefore proposed a simplification based on assuming $\avg{b_1\mu_1|\nu} \to \avg{b_1|\nu}\avg{\mu_1|\nu} = \hat b_1\avg{\mu_1|\nu}$, in which case the estimate for $b_{20}$ is (their equation 36)
\begin{align}
\delc^2 \hat b_{20}(\nu) &= \frac1{(1-\epc)^2}\bigg[\frac{\delc^2 \hat b_2}{(\Sc/\So)^2}\notag\\ 
&\ph{\bigg[\mu_2\bigg]}
- 2\epc\left(\frac{\delc \hat b_1}{(\Sc/\So)}\avg{\mu_1|\nu} - \avg{\mu_1^2|\nu}\right)\notag\\
&\ph{(1-\epc)^2\bigg[2\epc\mu_1\lam_1\bigg]}
- \epc(2-\epc)\avg{\mu_2|\nu} \bigg]\,.
\label{dc2b20-recon}
\end{align}
We will make this assumption in what follows, and leave a more detailed analysis of correlations such as $\avg{b_1\mu_1|\nu}$ to future work. 

\subsection{Measurements of $b_1$ and $b_2$}
\label{b1b2}

\begin{figure*}
 \centering
 \includegraphics[width=0.9\hsize]{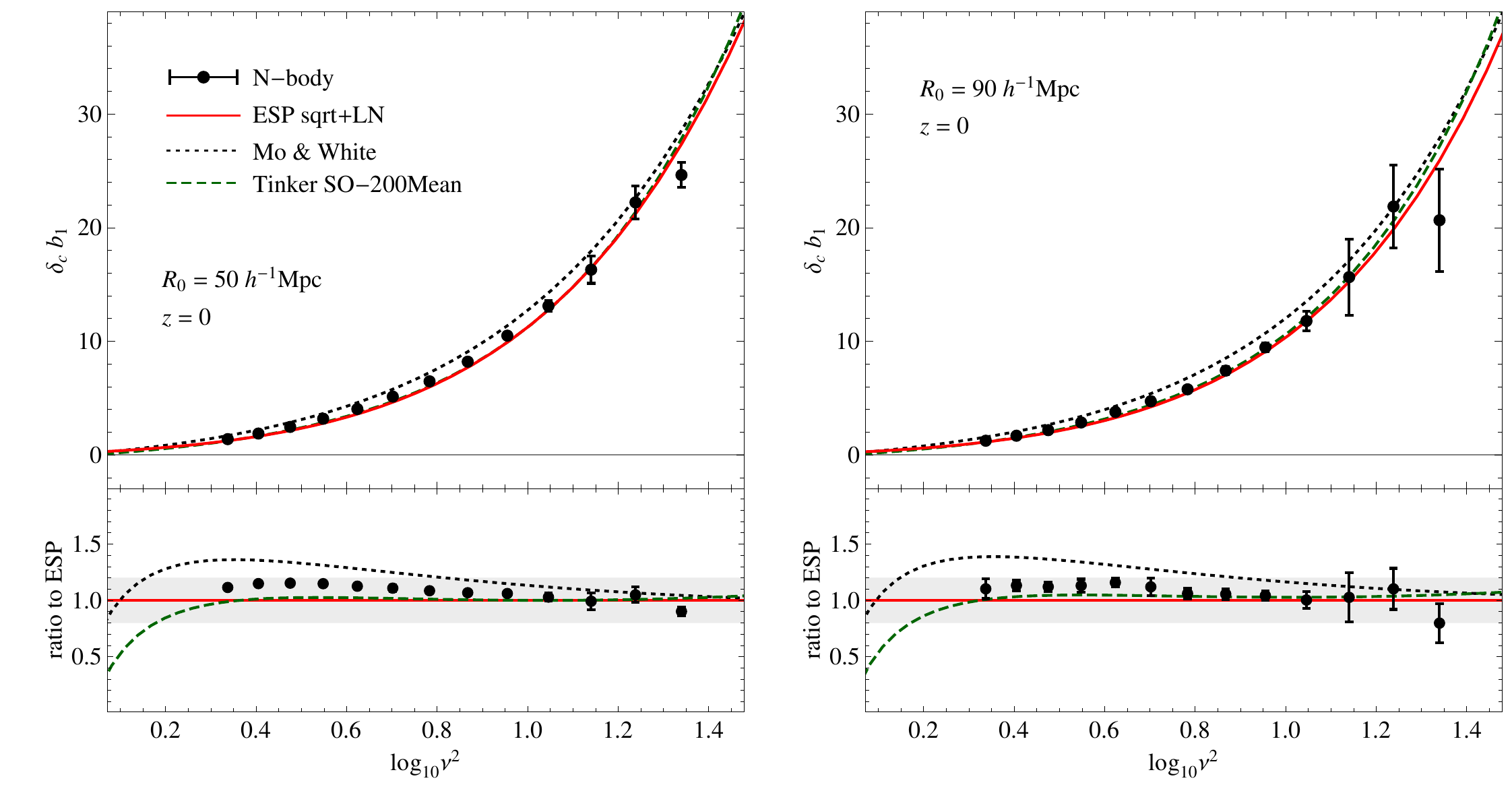}
 \caption{Measurements of the linear halo bias $b_1$ in the $N$-body simulations (points with error bars; see text for details). The two panels show measurements at two different smoothing scales. The smooth curves show the corresponding theory predictions using excursion set peaks (solid red) and from \citet[][dotted black]{mw96}, and the fit to large scale Fourier-space measurements in $N$-body simulations from \citet[][dashed green]{t+10}. We multiplied the latter two functions with the factor $\Sc/\So$ (which slowly varies with mass and is $\sim1.4$ for $R_0=50h^{-1}$Mpc and $\sim1.35$ for $R_0=90h^{-1}$Mpc) to account for the mapping from Fourier to real space. The excursion set peaks prediction includes the effect of both $\Sc/\So$ as well as \epc, although the latter effect is quite small at these scales. Lower panels show the ratios with respect to the ESP prediction, with the gray bands indicating $20\%$ deviations.
 }
 \label{fig-b1-sim}
\end{figure*}

\begin{figure*}
 \centering
  \includegraphics[width=0.9\hsize]{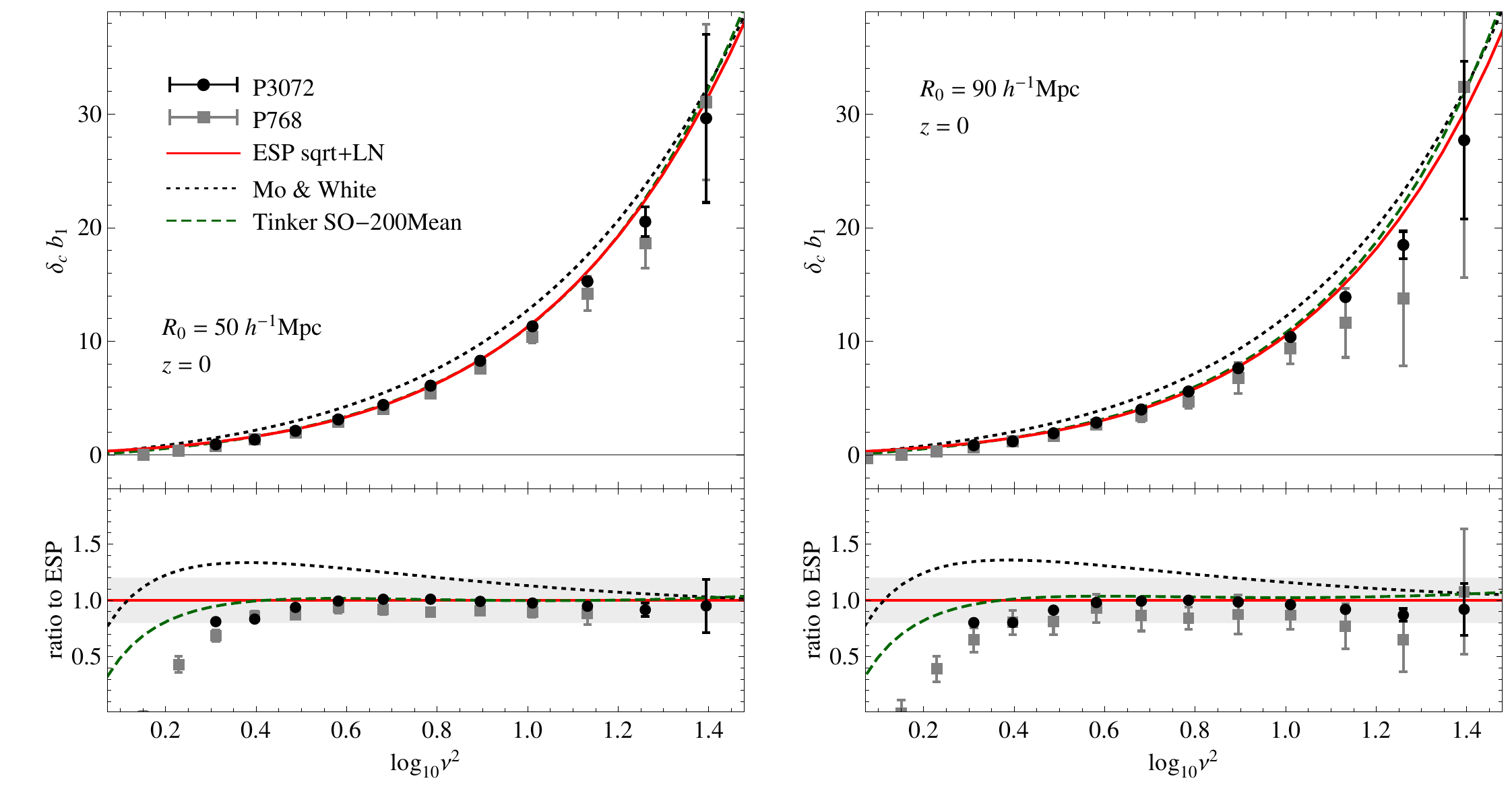}
 \caption{Same as \fig{fig-b1-sim}, now using measurements in \pino. The black circles and gray squares show measurements in the larger and smaller box, respectively. The latter show clear signs of finite volume effects for the larger smoothing scale. 
 }
 \label{fig-b1}
\end{figure*}

We now present measurements for the linear and quadratic bias coefficients defined by \eqn{bn-msd} and measured in the $N$-body and \pino\ runs described in section~\ref{sims}. These measurements, unlike the reconstruction algorithm which we implement in the next subsection, are completely independent of any theoretical input, other than the recognition that weighting by Hermite polynomials (rather than by simple powers of the density field) is the most natural way to define bias factors in a Gaussian random field \citep{s88,mps12}. 

\figs{fig-b1-sim} and \ref{fig-b1} show the results for $\delc b_1$ as measured by summing \eqn{bn-msd} over the halos identified at $z=0$ in the $N$-body simulations and in \pino, respectively.  The two panels in each Figure correspond to two different smoothing scales $R_0$. \fig{fig-b1} shows two sets of measurements; the black circles are from the larger simulation box and the gray squares from the smaller box. 

For the $N$-body simulations, the circles in \fig{fig-b1-sim} show the mean over $6$ runs and the error bars are the scatter around this mean for each bin. 
Similarly, the squares in \fig{fig-b1} show the mean over $9$ runs, and the error bars the scatter around this mean, for the smaller \pino\ box. 
For the larger \pino\ box, we had a single run, and the corresponding error bars in \figs{fig-b1-sim} and \ref{fig-b1} reflect Poisson errors for each mass bin. A comparison of the error bars shows that, at small masses, the Poisson errors are very likely underestimating the scatter in any given bin. 

These measurements are susceptible to at least two systematic finite volume effects; the first affects smoothing scales that are a significant fraction of the box size (e.g., $R_0=90h^{-1}$Mpc in the $768h^{-1}$Mpc \pino\ box), and the second affects \emph{mass bins} whose Lagrangian radius is comparable to the smoothing scale (e.g., $\log_{10}\nu^2(m,z=0)=1.4$ or $m\sim4.5\times10^{15}h^{-1}\Msun$ which has $R\sim15h^{-1}$Mpc, with $R_0=30h^{-1}$Mpc). Our choice of $R_0=50h^{-1}$Mpc tries to minimize these effects, while $R_0=90h^{-1}$Mpc highlights the first one. 

The solid red curves in \figs{fig-b1-sim} and \ref{fig-b1} are the corresponding predictions of the ESP framework at each smoothing scale \citep[equation 29 of][]{psd13}, for the respective cosmology. The dashed green lines in each panel are the Lagrangian bias from the fit presented by \citet{t+10}. To be consistent in comparing with our real space measurements, we multiplied this Fourier-space fit with a factor $\Sc/\So$ at each smoothing scale. The dotted black curves are the standard spherical collapse prediction $\nu^2-1$ \citep{mw96}, which we also multiplied by $\Sc/\So$.  
This is necessary because, although the Mo-White calculation is based on the excursion set approach, it implicitly assumes that the smoothing window with which $\delta_0$ is defined is a TopHat in $k$-space, for which $\Sc/\So = 1$, whereas our measurements (along with essentially all other real-space measurements) use a real space TopHat, a point first made by \cite{ps12a}.

The careful reader will have noticed that although the dashed and dotted curves are almost the same in the two panels (they differ only because the multiplicative factor $\Sc/\So$ is about 5\% different), the shape of the ESP prediction changes slightly.  This is due to the scale dependence introduced by \epc, although this effect is much smaller than the size of the error bars on the measurements. We see that the measurements of $b_1$ agree very well with the ESP prediction. The small systematic differences between the black circles and the red solid lines in each of the Figures can be traced back entirely to the respective mass functions; the ESP mass function slightly underpredicts the masses of the \pino\ halos and slightly overpedicts those in the $N$-body simulation (c.f. \fig{fig-mf}), which is consistent with the trends seen in \figs{fig-b1-sim} and \ref{fig-b1}.

\begin{figure*}
 \centering
   \includegraphics[width=0.9\hsize]{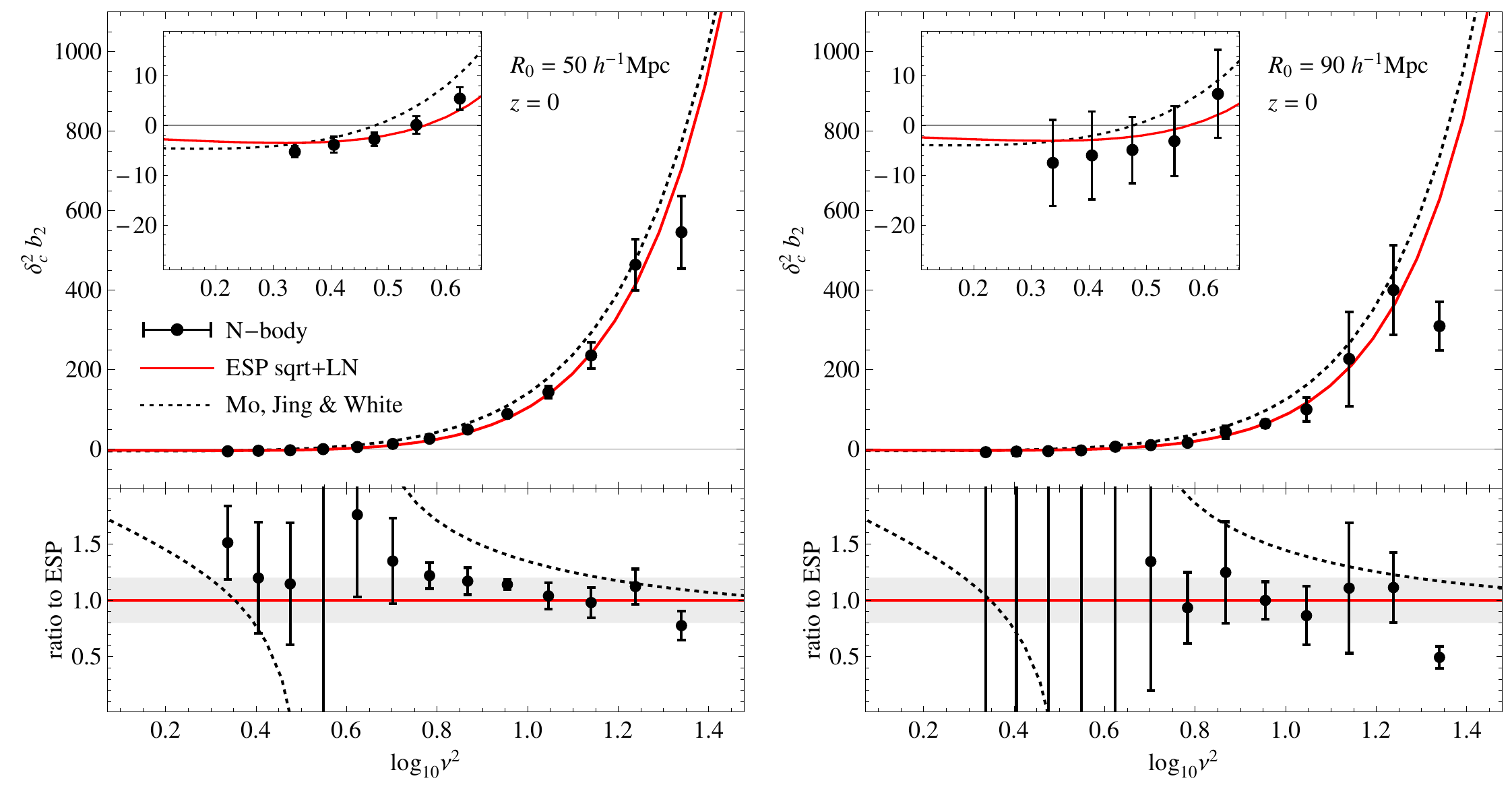}
 \caption{Measurements of the quadratic halo bias $b_2$ in the $N$-body simulations (points with error bars; see text for details). The left and right panels show measurements at two different smoothing scales (the same as in \fig{fig-b1-sim}). The smooth curves show the corresponding theory predictions using excursion set peaks (solid red) and from \citet[][dotted black]{mjw97}, the latter being multiplied with the factor $(\Sc/\So)^2$ to account for the mapping from Fourier to real space. In this case there are no prior $N$-body simulation results to compare with. Insets show zoomed-in views of the lowest masses. Lower panels show the ratios with respect to the ESP prediction, with the gray bands indicating $20\%$ deviations.
 }
 \label{fig-b2-sim}
\end{figure*}

\begin{figure*}
 \centering
   \includegraphics[width=0.9\hsize]{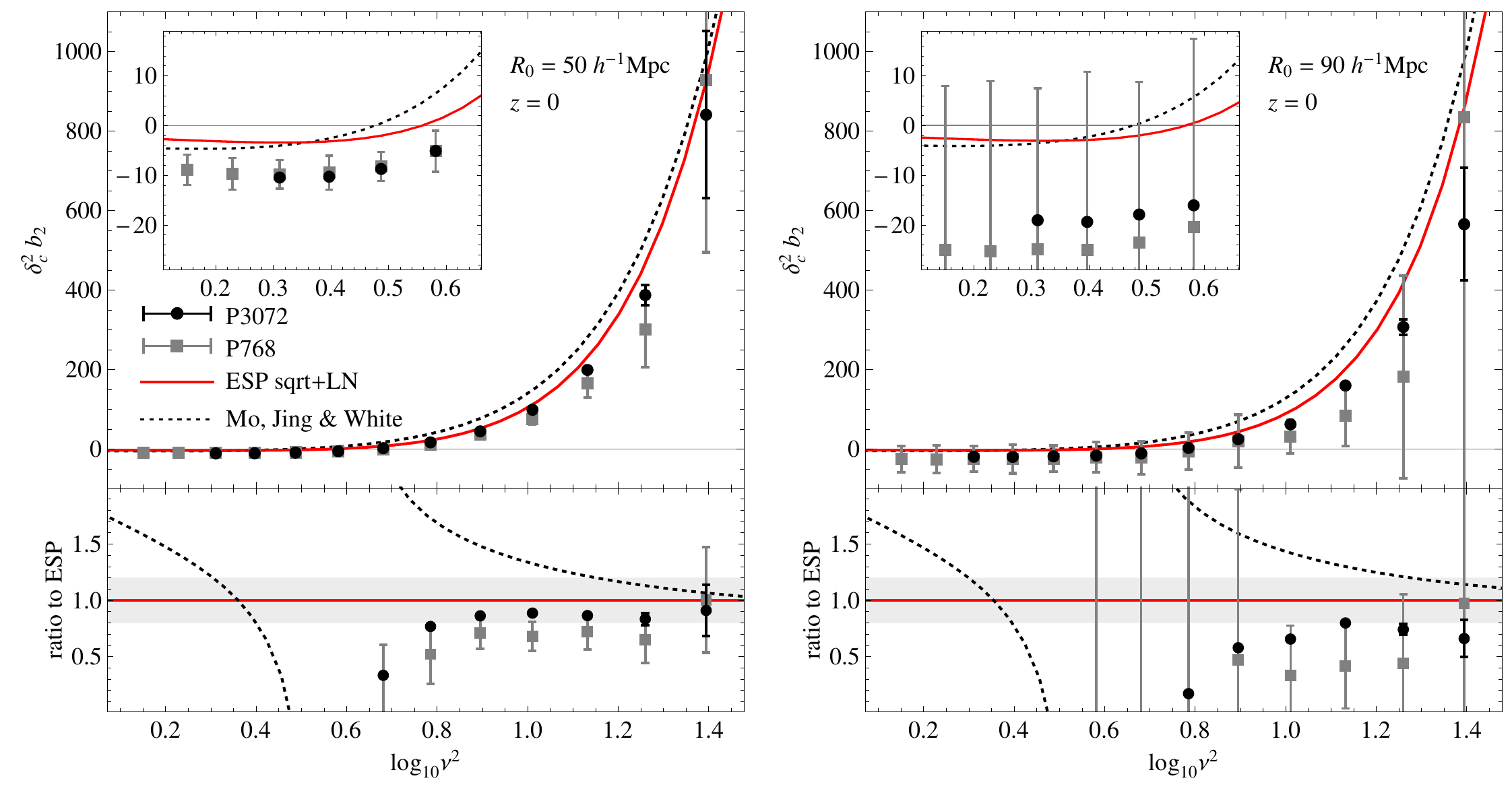}
 \caption{Same as \fig{fig-b2-sim}, now using measurements in \pino. The black circles and gray squares show measurements in the larger and smaller box, respectively. Finite volume effects are apparent at large masses for the measurements in the smaller simulation box.  
 }
 \label{fig-b2}
\end{figure*}

\figs{fig-b2-sim} and \ref{fig-b2} show the corresponding measurements with errors for the quadratic bias $\delc^2b_2$. 
The solid red curves in these Figures are the ESP prediction \cite[equation 30 of][]{psd13}. The dotted black curves are the spherical collapse prediction $\nu^2(\nu^2-3)$ \citep*{mjw97} multiplied this time by $(\Sc/\So)^2$ to account for the Fourier-to-real-space mapping. In this case there are no previous $N$-body results to compare with. In addition to the ratios with the ESP prediction in the lower panels, we also show an inset in each plot with a zoomed-in view of the smaller masses.

Once more we find good agreement of the $N$-body measurements with the ESP prediction (which also appears to be favored over the spherical collapse prediction). Note that the predicted (and measured) values go through zero close to $\log_{10}\nu^2\simeq0.6$. This is the first instance of a comparison between a parameter-free measurement of the quadratic bias coefficient $b_2$ with an analytical prediction, and the agreement we see is very encouraging. 
The measurements in \pino, while in reasonably good agreement with the ESP prediction, appear to show a systematic tendency to lie below the predictions, especially in the right panel of \fig{fig-b2}. 
We return to this issue later.

\subsection{Reconstructing $b_{10}$ and $b_{20}$}
\label{b10b20}

\begin{figure*}
 \centering
 \includegraphics[width=0.9\hsize]{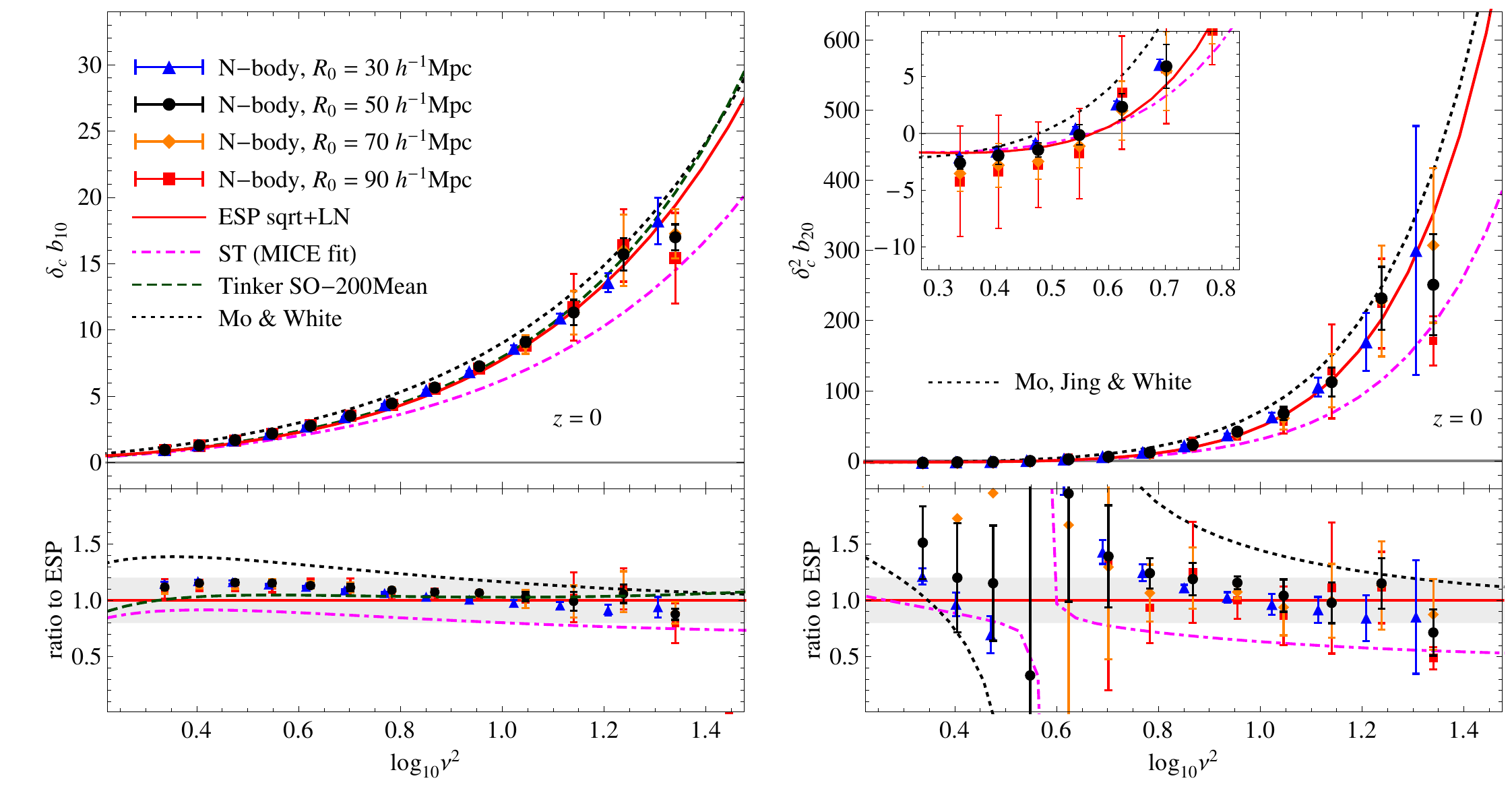}
 \caption{The reconstruction technique described in the text applied to the $N$-body measurements of linear and quadratic bias at $z=0$ at four different smoothing scales. \emph{Left:} Reconstructing $b_{10}$. 
The measurements are in excellent agreement with each other (indicating that the algorithm is working well)
and with the fit to Fourier-space measurements from \citet[][dashed green]{t+10} as well as the ESP prediction (solid red) apart from a minor trend caused by the mass mismatch seen in \fig{fig-mf}. For comparison we also show the prediction from \citet[][dotted black]{mw96} and the peak-background split prediction using the ST fit to the MICE mass function (the ST mass function with $q=0.7$, $p=0.26$ in equation~\ref{pbg} with $n=1$; dot-dashed magenta). \emph{Right:} Reconstructing $b_{20}$. 
The measurements again agree well with each other, although they are noisier than those of $b_{10}$. 
The smooth curves show the prediction of ESP (solid red), \citet[][dotted black]{mjw97} and the peak-background split prediction from the ST fit to the MICE mass function (dot-dashed magenta, the ST mass function with $q=0.7$, $p=0.26$ in equation~\ref{pbg} with $n=2$). The inset shows a zoom-in of the lowest masses. Note that the predicted values and measurements go through zero close to $\log_{10}\nu^2\simeq0.6$.
Lower panels in each case show the ratios with respect to the ESP prediction, with the gray bands indicating $20\%$ deviations.
 }
 \label{fig-b10b20-sim}
\end{figure*}

\begin{figure*}
 \centering
  \includegraphics[width=0.9\hsize]{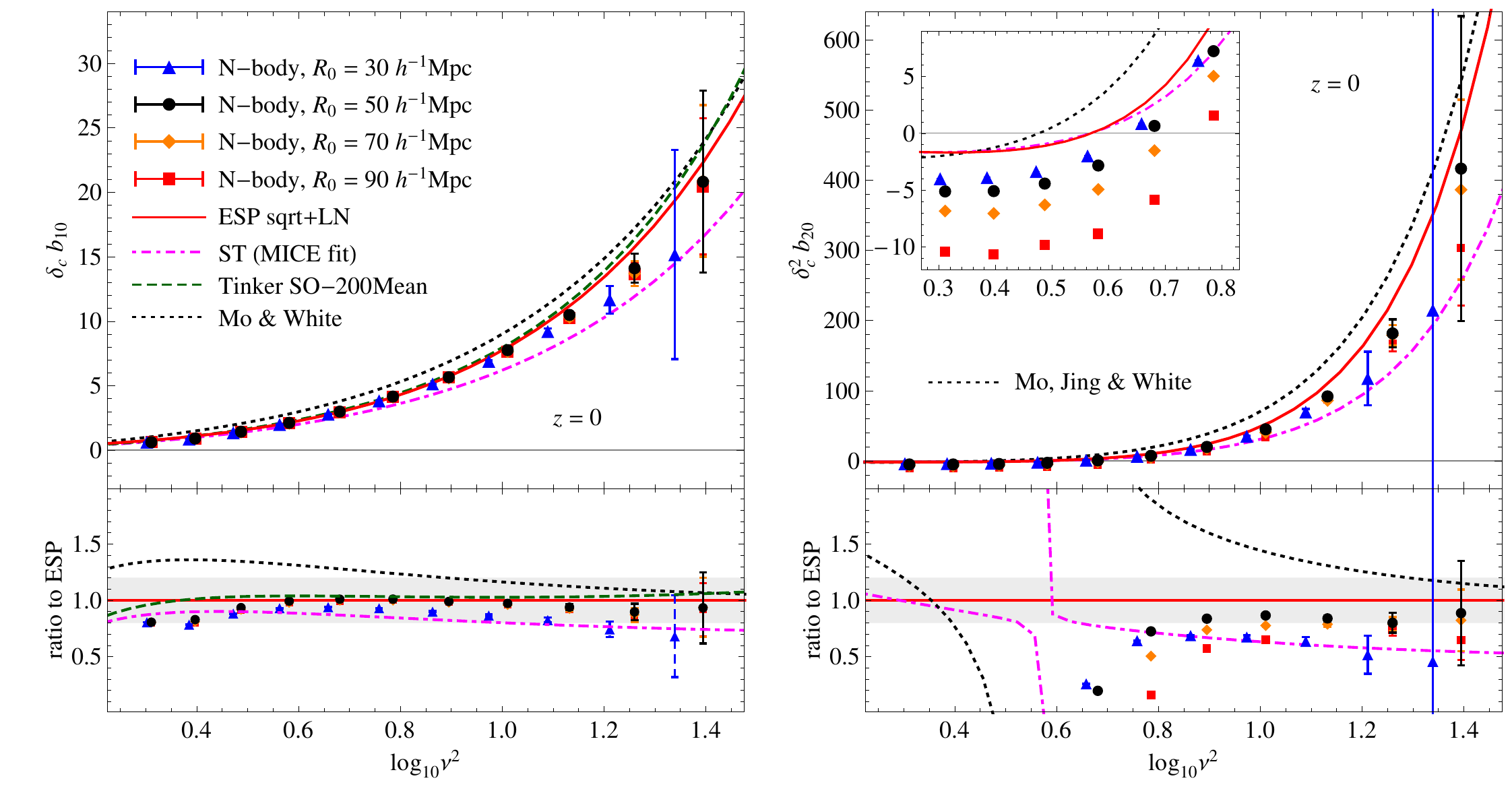}
 \caption{Same as \fig{fig-b10b20-sim}, now for the $3072\Mpc$ \pino\ box at $z=0$. 
In the left panel, the $b_{10}$ measurements 
are in excellent agreement with each other
(with some hints of systematic volume effects at $R_0=30\Mpc$)
and with the fit to Fourier-space measurements from \citet[][dashed green]{t+10} as well as the ESP prediction (solid red).
In the right panel, the $b_{20}$ measurements show a systematic decrease as $R_0$ increases.  
 }
 \label{fig-b10b20}
\end{figure*}

With the measurements of $b_1$ and $b_2$ in hand, we can apply \eqns{dcb10-recon} and~\eqref{dc2b20-recon} to reconstruct $b_{10}$ and $b_{20}$, and this requires some theoretical input. 

Firstly we must compute the functions $\Sc/\So$ and $\epc$ which are just integrals over the linear power spectrum (equations \ref{S0} and~\ref{Scepc}, see also footnote~\ref{fn1}). A more significant input is the value of the functions $\avg{\beta|\nu}$ and $\avg{\beta^2|\nu}$ which feed into the algorithm definition through \eqn{mu1mu2-avg}. As mentioned earlier, we evaluate these using the ESP prediction $f_{\rm ESP}(\nu|\beta)$ for the mass fraction at fixed $\beta$ and a Lognormal distribution for $\beta$ with mean $0.5$ and variance $0.25$, the same values used by \citet{psd13} which were derived from matching to the $N$-body results of \citet{rktz09}. 

The appearance of $\avg{\beta^j|\nu}$ which depends on the ESP mass function means that one cannot interpret the reconstructed $\hat b_{n0}$ as clean tests of the ESP formalism.
The original reconstruction prescription of \citet{mps12} only involved powers of $\nu$ (equation~\ref{bnr-mps12}) rather than integrals over some stochastic variable such as $\beta$, so one might argue that their prescription was in some sense cleaner and unaffected by choices regarding, e.g., the choice of distribution $p(\beta)$. This is misleading, however, since that prescription explicitly assumed a constant deterministic barrier which has been shown by \citet{psd13} to yield a poor description of the halo mass function, which is better described instead by the ingredients discussed above. 

The strength of our approach lies in the following. First, our reconstruction below of $b_{10}$, while model-dependent, agrees very well with the fit presented by \citet{t+10} to large scale \emph{Fourier}-space measurements in $N$-body simulations. Moreover, this model-dependence is almost entirely driven by the need to describe the mass function accurately \citep{psd13}, while the prediction for the bias, in a sense, comes for free. Secondly, our reconstruction of $b_{20}$ then makes \emph{no} additional assumptions regarding the underlying model, and is in this sense a parameter-free estimate of quadratic bias. The agreement between the scale-\emph{dependent} measurements of the previous subsection and the corresponding ESP predictions lends support to the expectation that this (albeit model-dependent) reconstruction scheme is correctly capturing the underlying physical processes that lead to halo bias.

\begin{figure*}
 \centering
 \includegraphics[width=0.9\hsize]{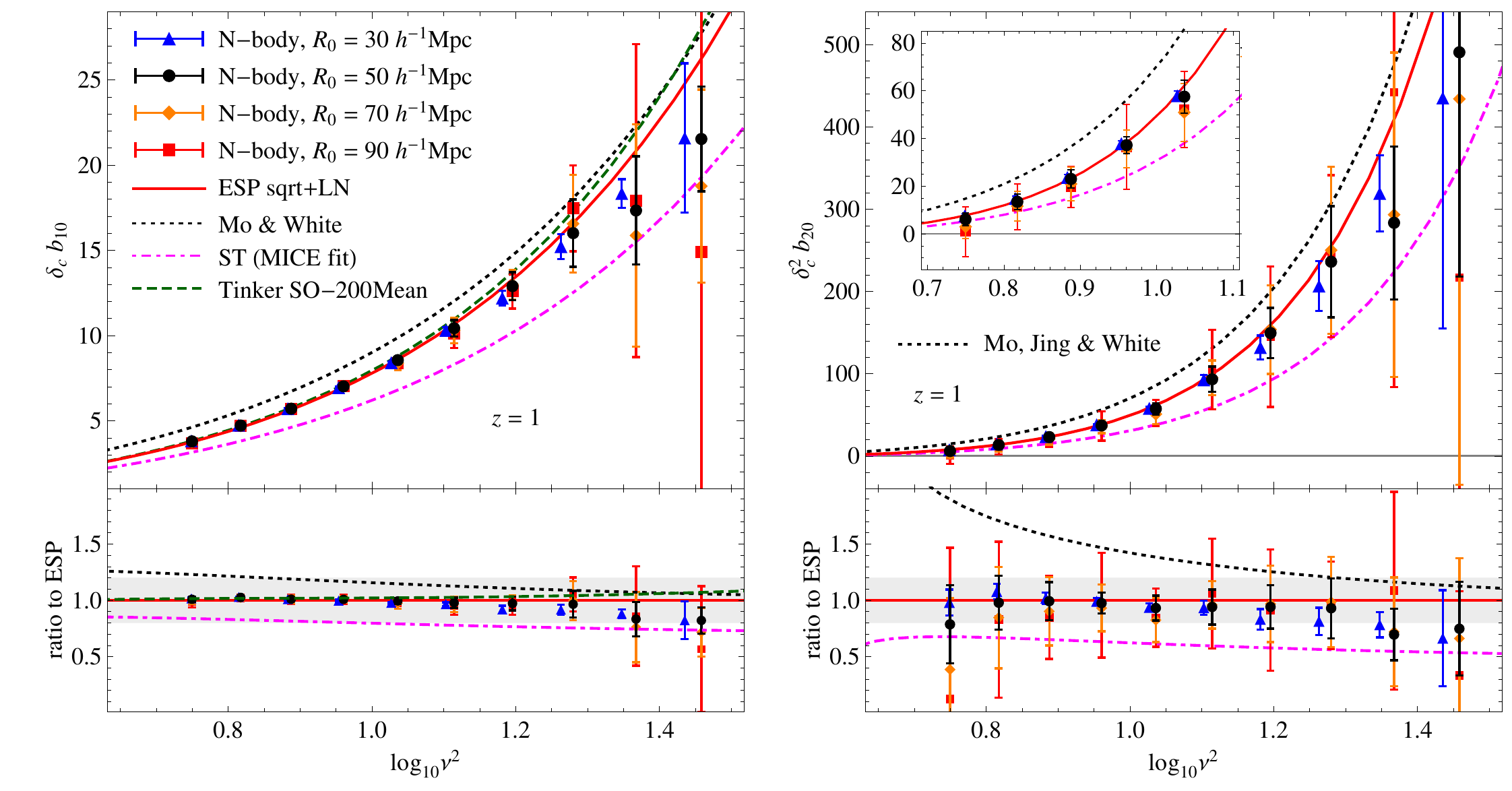}
 \caption{Same as \fig{fig-b10b20-sim}, now using halos identified at redshift $z=1$ in the $N$-body simulation. The measurements of $b_{10}$ are in excellent agreement with each other and with the Fourier-space fit from \citet{t+10} as well as the ESP prediction. Both $b_{10}$ and $b_{20}$ now display possible sample variance effects
at large masses ($\log_{10}\nu^2(m,z=1) \gtrsim 1.3$, or $m\gtrsim6\times10^{14}\Ms$).
See text for a discussion.
 }
 \label{fig-b10b20-z1p0-sim}
\end{figure*}

\begin{figure*}
 \centering
   \includegraphics[width=0.9\hsize]{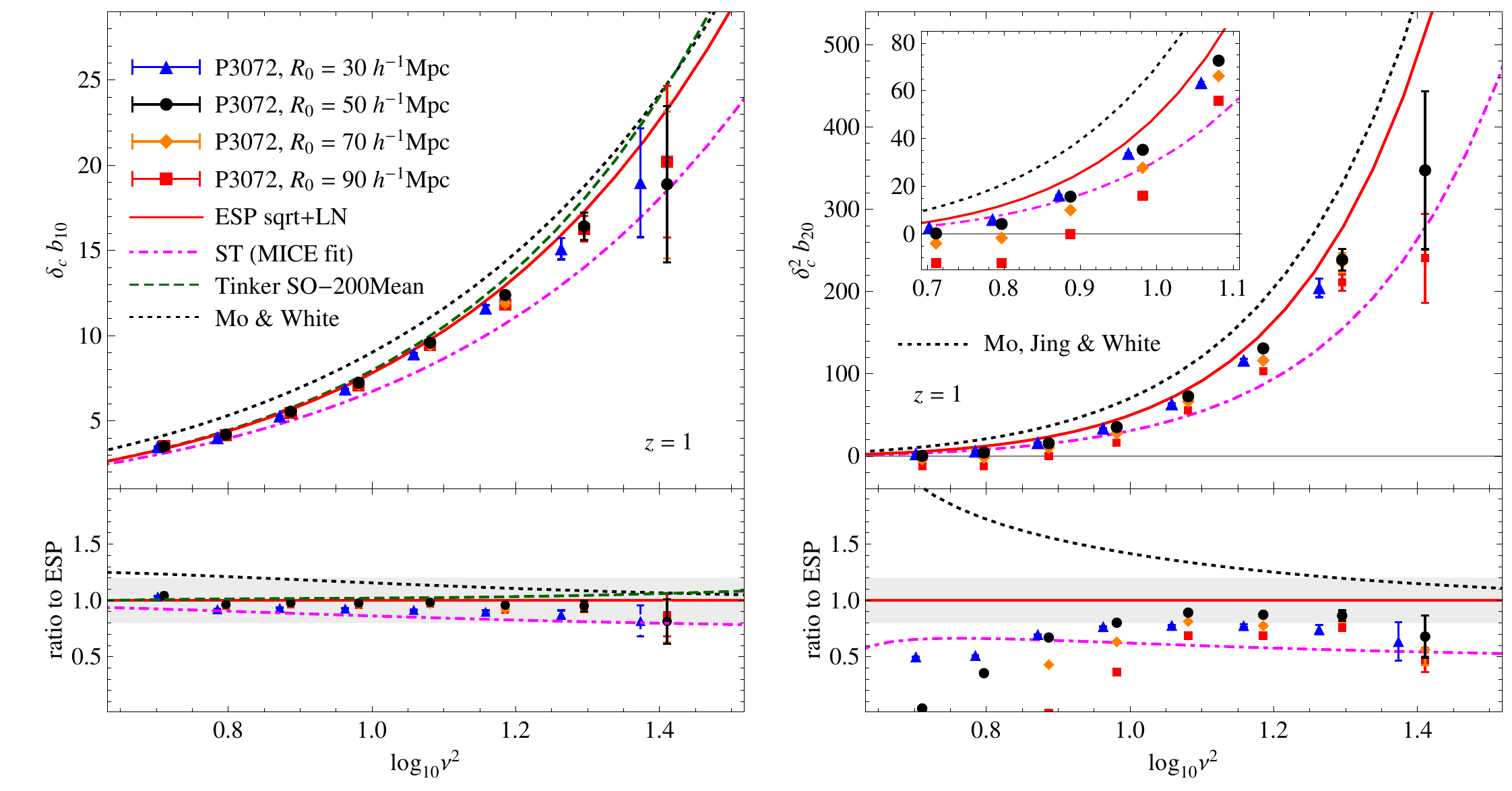}
 \caption{Same as \fig{fig-b10b20}, now using halos identified at redshift $z=1$ in the $3072\Mpc$ \pino\ box. 
See text for a discussion.
 }
 \label{fig-b10b20-z1p0}
\end{figure*}

The points with error bars in \fig{fig-b10b20-sim} show our reconstruction of $\delc b_{10}$ (left panel) and $\delc^2b_{20}$ (right panel) from measurements of $b_1$ and $b_2$ in the $N$-body simulations, for halos identified at $z=0$, on four different smoothing scales (two of which are the same as in the previous subsection). The errors were calculated by propagating the errors (i.e., scatter around the mean) on $b_1$ and $b_2$. \fig{fig-b10b20} shows the corresponding reconstructions in the $3072\Mpc$ \pino\ box; in this case the errors were computed by propagating Poisson errors.

We see in the left panel of \fig{fig-b10b20-sim} that the reconstruction of $b_{10}$ at all scales gives nearly identical results (indicating that the algorithm is working well in removing the scale dependence).
The reconstructed values are also in good agreement with the fit presented by \citet[][dashed green]{t+10} and the ESP prediction \citep[solid red; equation 29 of][with $\epc\to0$ and $\Sc/\So\to1$]{psd13}, apart from a minor trend caused by the mass mismatch seen in \fig{fig-mf}. For comparison, we also show the spherical collapse prediction $\nu^2-1$ \citep[][dotted black]{mw96} and the peak-background split prediction associated with the ST fit to the MICE mass function (the ST mass function in equation~\ref{pbg} with $n=1$; dot-dashed magenta).  

The reconstruction of $b_{20}$ in the right panel of \fig{fig-b10b20-sim} shows similar behaviour, and the measurements at all scales are in reasonable agreement with the ESP prediction \citep[solid red; equation 30 of][with $\epc\to0$ and $\Sc/\So\to1$]{psd13}. For comparison, we also show the spherical collapse prediction $\nu^2(\nu^2-3)$ \citep[][dotted black]{mw96} and 
the peak-background split prediction from the ST fit to the MICE mass function \citep[dot-dashed magenta, the ST mass function in equation~\ref{pbg} with $n=2$; see also][]{sshj01}.
In addition to the ratio with ESP predictions in the lower panels, the $b_{20}$ plot also shows an inset with a zoomed-in view of the lowest masses. Note that the predictions and measurements go through zero close to $\log_{10}\nu^2\simeq0.6$.

Correspondingly, the measurements of $b_{10}$ at $z=0$ in \pino\ (left panel of \fig{fig-b10b20}) are also in excellent agreement with each other and with the ESP prediction.
In this case, the reconstruction of $b_{20}$ in the right panel, while in reasonable agreement across scales, shows a systematic tendency towards lower values as $R_0$ is increased. We have noticed similar trends to a lesser extent in our individual $N$-body realisations as well (not displayed); these trends are the main reason that the scatter in the $b_{20}$ measurements in the right panel of \fig{fig-b10b20-sim} increases with smoothing scale $R_0$. 
The trend in the right panel of \fig{fig-b10b20} could therefore be due to sample variance. 

\figs{fig-b10b20-z1p0-sim} and \ref{fig-b10b20-z1p0} have a format identical to \figs{fig-b10b20-sim} and \ref{fig-b10b20}, but show results obtained using halos identified at $z=1$ in the $N$-body simulations (\fig{fig-b10b20-z1p0-sim}) and in the larger \pino\ box (\fig{fig-b10b20-z1p0}). Once again, the measurements of $b_{10}$ at different smoothing scales are in excellent agreement with each other 
and with the Fourier-space fit from \citet{t+10} as well as the ESP prediction. 
The $b_{20}$ measurements again show a systematic drift in the single \pino\ realisation and a large realisation-to-realisation scatter in the $N$-body results.
Additionally, there seems to be a systematic turnover at high masses ($\log_{10}\nu^2(m,z=1)\gtrsim1.3$ or $m\gtrsim6\times10^{14}\Ms$) in both $b_{10}$ and $b_{20}$ at all smoothing scales, which is more pronounced in the $N$-body simulation than in \pino. 
Since the \pino\ box has more than $8$ times the volume of our individual $N$-body realisations, this is also likely to be a sample variance effect.
We discuss other possible sources for these trends in Section \ref{discuss} below.

\section{Discussion}
\label{discuss}

The scale-dependence and nonlinearity of halo bias is a potential source of systematic effects for upcoming galaxy surveys, which can degrade the measurements of cosmological parameters. Current strategies for dealing with these effects mainly rely on using parametrized functions inspired by fits to $N$-body simulations \citep{t+05,psp12}. In principle, though, such effects can be modelled using analytical tools such as the excursion set formalism. While such models predict the linear, scale-independent part of halo bias with reasonable accuracy, they have not fared well until now in predicting these ``beyond linear'' effects.

We have presented real-space measurements of Lagrangian halo bias at linear and quadratic order using a recently  proposed technique \citep{mps12,psd13} which exploits the correlation between the locations of halo formation in the initial conditions and the large scale initial environment. The ``observables'' in this measurement -- namely, the scale dependent bias coefficients $b_n$ (equation~\ref{bn-msd}) -- can be estimated using simple measurements in the initial conditions of a simulation, and
are direct predictions of the excursion set peaks (ESP) formalism of \citet{psd13}. We find that our measurements in $N$-body simulations (\figs{fig-b1-sim} and \ref{fig-b2-sim}) and in the \pino\ algorithm (\figs{fig-b1} and \ref{fig-b2}) are in very good agreement with expectations using ESP.

Further, the ESP formalism also shows how these scale-dependent measurements of $b_n$ can be converted into scale-\emph{independent} estimates of the peak-background split parameters $b_{n0}$ defined by \eqn{pbg}. We find good agreement between our estimates of $b_{10}$ in \figs{fig-b10b20-sim}, \ref{fig-b10b20}, \ref{fig-b10b20-z1p0-sim} and \ref{fig-b10b20-z1p0} and the fit to large scale \emph{Fourier}-space linear bias in $N$-body simulations presented by \citet{t+10}, which represents a success of the ESP formalism. Furthermore, the reconstruction of $b_{n0}$ utilizes no additional assumptions other than those required to obtain a good description of the mass function; in this sense, we have presented a parameter-free prescription for estimating nonlinear large scale (i.e. small $k$ or large $R_0$) bias from intermediate scale measurements (we used $R_0=30,50,70,90\Mpc$).

Our analysis and results raise a number of interesting issues that must be addressed to assess and utilize the full potential of our approach:
\begin{itemize}
\item We note that our estimates of $b_{10}$ and $b_{20}$ show departures from the theoretical predictions at the largest halo masses; these are especially pronounced in the $N$-body results at $z=1$ (\fig{fig-b10b20-z1p0-sim}). As discussed in the text, a likely cause for these is sample variance. 
Additionally, at small smoothing scales like $R_0=30\Mpc$ the trends could be partly due to the fact that the largest halos have Lagrangian sizes that approach this size. 
There is also the possibility that nonlocality induced by effects such as those studied by \citet{scs13} could be contaminating our reconstruction algorithm at high masses.
It will be interesting to check whether such nonlocal, nonspherical effects can be isolated from the recovery of the $b_{n0}$, e.g., using the rotational-invariance motivated orthogonal polynomials associated with peak shapes \citep{gpp12,d13} and the tidal field \citep{scs13}. This should require straightforward extensions of our technique to measuring quantities other than the density.
\item Another important question is whether our technique, which works at finite scales, can become a realistic, cost-effective method for measuring bias in the \emph{late-time}, gravitationally evolved Eulerian field. We have presented results using smoothing scales $R_0\geq30\Mpc$, which are large compared with, e.g., those used by \citet{abl08} in their Eulerian analysis that recovered bias coefficients up to $4^{\rm th}$ order. The reason for having these large Lagrangian smoothing scales is simply that the most massive protohalos we work with have Lagrangian sizes that can approach $\sim15$-$20\Mpc$. Halos in the Eulerian field are much more compact, so in principle our technique should be applicable at much smaller scales in this case. A bigger issue is the fact that the Eulerian field is weakly non-Gaussian, so using Hermite polynomials as we currently do may not be the best approach. Although one can find natural generalisations of the Hermite polynomials appropriate for non-Gaussian fields \citep[e.g., see Appendix B2 of][]{mps12}, the question of which basis set to use in extracting nonlinear Eulerian bias is still not fully settled. 
\item Our original goal included a measurement of the $k$-dependence of linear bias as implied by the presence of the quantity \epc\ in the excursion set prediction in \eqn{bn-expand}; however, this effect is too small to be reliably seen given our present error bars. 
For scale-dependent bias in the \emph{Eulerian} field, there is also the related question of whether triaxiality of halos could be a source of significant systematics for our reconstruction technique. The Eulerian field will also be affected by gravitationally induced nonlocality \citep{css12} which would have to be accounted for.
\end{itemize}

And finally, it will be very interesting to assess to what extent a self-consistent estimate of nonlinear and nonlocal halo bias in galaxy surveys can improve the recovery of cosmological parameters. We leave a detailed exploration of all these issues to future work.

\section*{Acknowledgments}
We thank Roman Scoccimarro for collaboration in the initial stages of this work, and both him and Marcello Musso for useful discussions.  KCC and VD acknowledge support by the Swiss National Science Foundation.  RKS is supported in part by NSF-AST 0908241.  

\label{lastpage}


\begin{thebibliography}{}
 \bibitem[\protect\citeauthoryear{Angulo, Baugh \& Lacey}{Angulo et al.}{2008}]{abl08} 
   Angulo R.~E., Baugh C., Lacey C., 2008, MNRAS, 387, 921
 \bibitem[\protect\citeauthoryear{Baldauf et al.}{2012}]{bsdm12} 
  Baldauf T., Seljak U., Desjacques V., McDonald P., 2012, PRD, 86, 083540
 \bibitem[\protect\citeauthoryear{Blake et al.}{2011}]{b+11} 
   Blake C. et al., 2011, MNRAS, 418, 1707
 \bibitem[\protect\citeauthoryear{Bryan \& Norman}{1998}]{bn98} 
   Bryan G.~L., Norman M.~L., 1998, ApJ, 495, 80
 \bibitem[\protect\citeauthoryear{Chan \& Scoccimarro}{2012}]{cs12} 
   Chan K.~C., Scoccimarro R., 2012, PRD, 86, 103519
 \bibitem[\protect\citeauthoryear{Chan, Scoccimarro \& Sheth}{Chan et al.}{2012}]{css12} 
   Chan K.~C., Scoccimarro R., Sheth R.~K., 2012, PRD, 85, 083509
 \bibitem[\protect\citeauthoryear{Crocce et al.}{2010}]{mice}
   Crocce M., Fosalba P., Castander F.~J., Gazta\~naga E., 2010, MNRAS, 403, 1353
 \bibitem[\protect\citeauthoryear{Crocce, Pueblas \& Scoccimarro}{Crocce et al.}{2006}]{cpr06}
   Crocce M., Pueblas S., Scoccimarro R., 2006, MNRAS, 373, 369
 \bibitem[\protect\citeauthoryear{Desjacques et al.}{2010}]{dcss10} 
   Desjacques V., Crocce M., Scoccimarro R., Sheth R.~K., 2010, PRD, 82, 103529
 \bibitem[\protect\citeauthoryear{Desjacques}{2013}]{d13} 
   Desjacques V., 2013, PRD, 87, 043505
 \bibitem[\protect\citeauthoryear{Despali, Tormen \& Sheth}{Despali et al.}{2013}]{dts13}
   Despali G., Tormen G., Sheth R.~K., 2013, MNRAS, 431, 1143
 \bibitem[\protect\citeauthoryear{Eke, Cole \& Frenk}{Eke et al.}{1996}]{ecf96}
   Eke V.~R., Cole S., Frenk C.~S., 1996, MNRAS, 282, 263
 \bibitem[\protect\citeauthoryear{Fry \& Gazta\~naga}{1993}]{fg93}
   Fry J.~N., Gazta\~naga E., 1993, ApJ, 413, 447
 \bibitem[\protect\citeauthoryear{Gay, Pichon, Pogosyan}{Gay et al.}{2012}]{gpp12} 
   Gay C., Pichon C., Pogosyan D., 2012, Phys Rev D., 85, 023011
 \bibitem[\protect\citeauthoryear{Gill, Knebe \& Gibson}{Gill et al.}{2004}]{gkg04}
   Gill S.~P.~D., Knebe A., Gibson B.~K., 2004, MNRAS, 351, 399
 \bibitem[\protect\citeauthoryear{Kaiser}{1984}]{k84}
   Kaiser N., 1984, ApJ, 284, L9
 \bibitem[\protect\citeauthoryear{Knollmann \& Knebe}{2009}]{kk09}
   Knollmann S.~R., Knebe A., 2009, ApJS, 182, 608
 \bibitem[\protect\citeauthoryear{Lewis, Challinor \& Lasenby}{Lewis et al.}{2000}]{camb}
   Lewis A., Challinor A., Lasenby A., 2000, ApJ, 538, 473
 \bibitem[\protect\citeauthoryear{Manera \& Gazta\~naga}{2012}]{mg11}
   Manera M., Gazta\~naga E., 2011, MNRAS, 415, 383
 \bibitem[\protect\citeauthoryear{Matsubara}{2012}]{m11} 
   Matsubara T., 2011, PRD, 83, 083518
 \bibitem[\protect\citeauthoryear{Mo, Jing \& White}{Mo et al.}{1997}]{mjw97} 
   Mo H.~J, Jing Y.~P., White S.~D.~M., 1997, MNRAS, 284, 189
 \bibitem[\protect\citeauthoryear{Mo \& White}{1996}]{mw96} 
   Mo H.~J., White S.~D.~M., 1996, MNRAS, 282, 347
 \bibitem[\protect\citeauthoryear{Monaco et al.}{2002}]{mo02} 
   Monaco P., Theuns T., Taffoni G., 2002, MNRAS, 331, 587
 \bibitem[\protect\citeauthoryear{Monaco et al.}{2013}]{mo13} 
   Monaco P., Sefusatti E., Borgani S., Crocce M., Fosalba P., Sheth R.~K., Theuns T., {\em accepted  by} MNRAS
 \bibitem[\protect\citeauthoryear{Musso, Paranjape \& Sheth}{Musso et al.}{2012}]{mps12} 
   Musso M., Paranjape A., Sheth R.~K., 2012, MNRAS, 427, 3145 
 \bibitem[\protect\citeauthoryear{Musso \& Sheth}{2012}]{ms12} 
   Musso M., Sheth R.~K., 2012, MNRAS, 423, 102
 \bibitem[\protect\citeauthoryear{Paranjape \& Sheth}{2012a}]{ps12a} 
   Paranjape A., Sheth R.~K., 2012a, MNRAS, 419, 132
 \bibitem[\protect\citeauthoryear{Paranjape \& Sheth}{2012b}]{ps12b} 
   Paranjape A., Sheth R.~K., 2012b, MNRAS, 426, 2789
 \bibitem[\protect\citeauthoryear{Paranjape, Sheth \& Desjacques}{Paranjape et al.}{2013}]{psd13} 
   Paranjape A., Sheth R.~K., Desjacques V., 2013, MNRAS, 431, 1503
 \bibitem[\protect\citeauthoryear{Pollack, Smith \& Porciani}{Pollack et al.}{2012}]{psp12} 
   Pollack J.~E., Smith R.~E., Porciani C., 2012, MNRAS, 420, 3469
 \bibitem[\protect\citeauthoryear{Robertson et al.}{2009}]{rktz09} 
   Robertson B.~E., Kravtsov A.~V., Tinker J., Zentner A.~R, 2009, ApJ, 696, 636
 \bibitem[\protect\citeauthoryear{S\'anchez et al.}{2012}]{s+12} 
   S\'anchez A. G. et al., 2012, MNRAS, 425, 415
 \bibitem[\protect\citeauthoryear{Scoccimarro et al.}{2001}]{sshj01} 
   Scoccimarro R., Sheth R.~K., Hui L., Jain B., 2001, ApJ, 546, 20
 \bibitem[\protect\citeauthoryear{Sheth \& Tormen}{1999}]{st99} 
   Sheth R.~K., Tormen G., 1999, MNRAS, 308, 119
 \bibitem[\protect\citeauthoryear{Sheth, Chan \& Scoccimarro}{Sheth et al.}{2013}]{scs13} 
   Sheth, R.~K., Chan, K.~C., Scoccimarro, R., 2013, PRD, 87, 083002
 \bibitem[\protect\citeauthoryear{Springel}{2005}]{gadget05} 
   Springel V., 2005, MNRAS, 364, 1105 
 \bibitem[\protect\citeauthoryear{Szalay}{1988}]{s88} 
   Szalay A., 1988, ApJ, 333, 21 
 \bibitem[\protect\citeauthoryear{Tinker et al.}{2008}]{t+08} 
   Tinker J.~L., Kravtsov A.~V., Klypin A., Abazajian K., Warren M.~S., 
   Yepes G., Gottl{\"o}ber S., Holz D.~E., 2008, ApJ, 688, 709
 \bibitem[\protect\citeauthoryear{Tinker et al.}{2010}]{t+10} 
   Tinker J.~L., Robertson B.~E., Kravtsov A.~V., Klypin A., 
    Warren M.~S., Yepes G., Gottl{\"o}ber S., 2010, ApJ, 724, 878
 \bibitem[\protect\citeauthoryear{Tinker et al.}{2005}]{t+05} 
   Tinker, J. L., Weinberg, D. H., Zheng, Z., Zehavi, I. 2005, ApJ, 631, 41
 \bibitem[\protect\citeauthoryear{Warren et al.}{2006}]{wa06} 
    Warren M. S., Abazajian K., Holz D. E., Teodoro L., 2006, ApJ, 646, 881
\end{thebibliography}
\end{document}